# GLEAKE: Global and Local Embedding Automatic Keyphrase Extraction

Javad Rafiei Asl[1] and Juan M. Banda[2]

**Abstract**—Automated methods for granular categorization of large corpora of text documents have become increasingly more important with the rate scientific, news, medical, and web documents are growing in the last few years. Automatic keyphrase extraction (AKE) aims to automatically detect a small set of single or multi-words from within a single textual document which capture the main topics of the document. AKE plays an important role in various NLP and information retrieval tasks such as document summarization and categorization, full-text indexing and article recommendation. Due to the lack of sufficient human-labeled data in different textual contents, supervised learning approaches are not ideal for automatic detection of keyphrases from the content of textual bodies. With the state-of-the-art advances in text embedding techniques, NLP researchers have focused on developing unsupervised methods to obtain meaningful insights from raw datasets. In this work, we introduce Global and Local Embedding Automatic Keyphrase Extractor (GLEAKE) for the task of AKE. GLEAKE utilizes single and multi-word embedding techniques to explore the syntactic and semantic aspects of the candidate phrases and then combines them into a series of embedding-based graphs. Moreover, GLEAKE applies network analysis techniques on each embedding-based graph to refine most significant phrases as a final set of keyphrases. We demonstrate the high performance of GLEAKE by evaluating its results on five standard AKE datasets from different domains and writing styles and by showing its superiority with regards to other state-of-the-art methods.

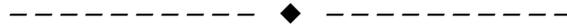

## 1 INTRODUCTION

Keyword or keyphrase extraction is related to selecting a concise set of single or multi-terms from within a text entity that characterize the main topics of its content. As a fundamental task, Automatic Keyphrase Extraction (AKE) plays an important role in many natural language processing and information retrieval projects; Hulth and Megyesi [1] combined automatically extracted keywords with full-text representation to improve the text categorization task. Ferrara et al. [2] used a keyphrase extraction module as the main component of a content-based recommendation system to improve the access to scientific digital libraries. Using keyphrases instead of n-grams have elevated two approaches to the top rank in cross-lingual text reuse detection [3]. Guan has applied keyphrase extraction techniques to extract key-terms in order to answer biomedical questions. These key-terms can be used in query-based summarization as core concepts of abstractive summarization to answer biomedical questions which are user understandable [4]. Additionally, AKE methods help scientific publishers to recommend articles to readers, highlight missing citations for authors, identify potential reviewers for manuscript submissions, and analyze research trends over time [5].

Due to the complexity of scientific documents and the considerable volume of documents being generated, the amount of time and effort to manually extract concise sets of keyphrases has become unfeasible. To mitigate this problem, researchers have recently assigned much attention to automatic keyphrase extraction (AKE) [5], [6]. Supervised AKE approaches need to be trained on large annotated corpora, which requires a considerable human effort and resources to generate manual keyphrases [7]. Additionally, supervised methods usually fail to generalize across multiple domains [8].

In this work, we introduce GLEAKE, an unsupervised method for AKE using a combination of local cues and semantic insights from candidate phrases. For each input document, the proposed method identifies a large set of relevant phrases as candidates using a customized noun-phrase pattern. After this initial step, a local word embedding model is trained on the input document and used by GLEAKE to assign a syntactic vector for each candidate phrase and the document. Our method then employs a single or multi-word embedding to infer semantic vectors for the candidates and the input document. All obtained local and global information from the previous steps is merged into a series of embedding-based graphs where the nodes are candidate phrases and the edges are determined using several weighting functions of syntactic and semantic similarities. Lastly, our method utilizes network analysis techniques to score and rank the candidate phrases and select top N candidates as the final set of keyphrases. We have evaluated GLEAKE performance by

1.) Investigating four significant hyper-parameters from the model's architecture.
2.) Examining the efficiency of the local and global information mapper separately and interpreting how GLEAKE outperform both of them.
3.) And finally comparing it with three baselines and

- *1. Office 752, Department of Computer Science, Georgia State University, P.O. Box 5060, Atlanta, GA 30302-5060. E-mail: jasl1@student.gsu.edu.*
- *2. Office 749, Department of Computer Science, Georgia State University, P.O. Box 5060, Atlanta, GA 30302-5060. E-mail: jbanda@gsu.edu.*

five state-of-the-art AKE methods.

The rest of the paper is organized as follows. In section 2, we would discuss AKE related works. Section 3 introduces our method and describes all its components elaborately. In section 4, we would concentrate on our experimental evaluation in order to determine the best hyperparameters and compare GLEAKE performance with other state-of-art AKE methods. Lastly, we will present our conclusion in section 5 and discuss useful directions for future work.

## 2 RELATED WORK

In recent years, supervised keyphrase extraction techniques have presented better efficiency than unsupervised techniques considering a specific domain of knowledge. However, the main problem with the supervised methods is their limitation to generalize well to other domains of content [8]. Therefore, due to being free from any specific context, unsupervised approaches demonstrate their superiority when they are applied to different domains. Nowadays with the advent of word embeddings, unsupervised methods represent their dominance on different fields of NLP and information retrieval [9], [10], [11]. In the following subsections, we briefly explain some of the state-of-the-art unsupervised AKE methods and also review a variant of multi-word embeddings that are applicable in AKE task.

### 2.1 Unsupervised Keyphrase Extraction

These approaches aim to investigate the intrinsic characteristics of keyphrases without utilizing any annotated dataset. Mihalcea et al. [12] proposed TextRank, a classical unsupervised method in which a directed graph is built considering words as nodes and word co-occurrences within the text document as links between nodes. Then an algorithm derived from Google's PageRank is applied to rank the words according to their importance. In Grineva et al. [13], the authors converted the input document into a word graph of semantic relationships between words. Several community detection techniques are employed to divide the graph into thematically cohesive communities. Finally, two important properties of the communities, density and informativeness, are applied on ranking and selecting communities that contain key-terms. Liu et al. [7] proposed Topical PageRank (TPR), which applied TextRank multiple times, each one specific to a topic of input document. The topics were induced by a Latent Dirichlet Allocation (LDA) model proposed by Blei [14] and thus the final set of keyphrases would cover all the main topics of the document according to their proportion in its content.

In the recent works, Kumar et al. [15] combines the statistical features obtained by weighted betweenness centrality and semantic expansion strategies obtained by using NPMI score to identify the candidate keyphrases. Haddoud and Mokhtari investigate the impact of candidate terms filtering through five POS tag sequence definitions of a noun phrase [16]. SemGraph [17] is the other unsupervised algorithm for extracting keyphrases based on a semantic relationship graph resulted from the combination of the co-occurrence graph and lexical knowledge of WordNet. Meng et al. use recurrent neural networks (RNN) to compress the semantic information in the given text into a dense vector [18]. Furthermore, the authors incorporate a copying mechanism to allow their model to find important parts based on positional information. Thus, the model can generate keyphrases based on an understanding of the text; at the same time, it does not lose important in-text information.

### 2.2 Sequence of Words Embeddings

Multi-word embeddings refer to a series of models that represent an arbitrary sequence of words as a fixed-length semantic vector in a continuous vector space. During recent years, several multi-word embedding techniques were introduced: The Skip-Thought model tries to predict the surrounding sentences of an encoded sentence, so the sentences with close semantic properties have similar vector representations [19]. The Doc2vec model [20] considers every paragraph as a token with a unique vector that averaged or concatenated with a fixed-length context of words inside the paragraph to predict the next word in the context. Mahata et al. [21] trained multi-word phrase embedding using the Fasttext[1] Framework. Sent2vec [22] embeds word alongside n-gram vectors, simultaneously trains their composition to produce a sentence vector. Cer et al. [23] presents two models for producing sentence embeddings: The first model constructs a sentence embedding, using encoding the sub-graph of the transformer architecture [24], where the sub-graph uses attention to compute context aware representations of words in a sentence to take into account both the ordering and identity of all the other words. The second encoding model makes use of a deep averaging network (DAN) [25], so that the input embeddings of the words and bi-grams are first averaged together and then passed through a feedforward deep neural network to produce sentence embeddings.

## 3 OUR APPROACH

As shown in Fig. 1, GLEAKE consists of several independent components that cooperate effectively to output a comprehensive set of high-quality keyphrases for each input document. In this section, we will elaborate each component specifically and also clarify how the information flows among the components to produce the high-quality keyphrases. Typically, an input document is passed from five main steps to produce a set of final keyphrases: At first, the document is preprocessed and then significant candidate keyphrases are extracted from its content. In the second step, we train a word embedding model on the input document and utilize it to obtain a local vector for each candidate phrase and the complete input document. At the next step, we employ a proper single or multi-word embedding model regarding the content type of the document to assign one global vector for each candidate and the input document. The fourth step, as the most crucial phase in our approach, integrates the local and global information from two previous steps into an embedding-based graph. Inside the graph, the candidates, as nodes are connected to each other according to similarities of the local and global vectors. In the final step, GLEAKE applies several social network centralities on the embedding-based network to assign a score for each candidate phrase, rank them, and finally select top N candidate phrases as the final keyphrases.

---
[1] https://fasttext.cc/

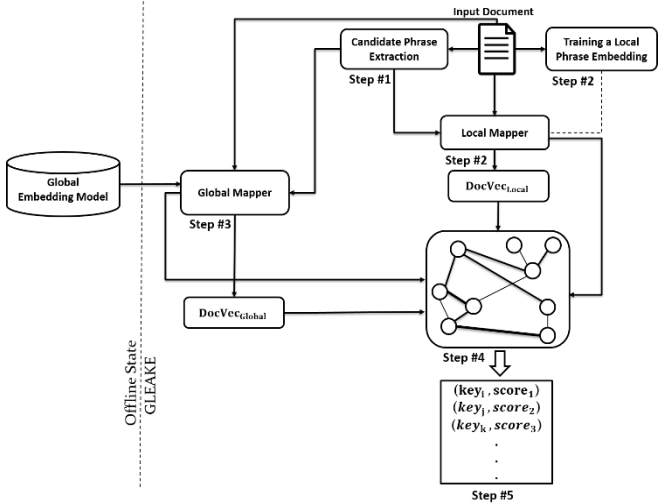

Fig. 1. GLEAKE architecture - Five main components.

In this work, we introduce a new concept in the NLP domain called embedding-based graph where the graph is completely constructed using the embedded vectors of phrases instead of using the co-occurrence data of words. The graph combines the local cues with the semantic concepts of early phrases to make it easier to retrieve the most significant phrases that capture the main ideas inside the document. As a feature, GLEAKE has the capability to readily replace any component of its architecture with a relevant alternative and improve its efficiency in various types of context. In this research, we have concentrated on scientific and news contents and demonstrate GLEAKE's high efficiency on five different AKE datasets. However, GLEAKE also contains alternative mapper components to extract the meaningful set of keyphrases from other domains of content, such as twitter, medical, and common content.

In the following subsections, we separately investigate each main component of GLEAKE and explain how these components interact with each other to produce a high-quality set of keyphrases for each input document.

### 3.1 Candidate Phrase Extraction

Before addressing the candidate phrase issue, GLEAKE applies simple preprocessing operations on the input content. These operations include: integrating all lines into one line, replacing multi-spaces and tab characters with single space characters, converting all letters to lowercase, and removing special characters and digits from the input content. These simple operations refine the content of the input document and make it more convenient to seek the keyphrases. The next step is extracting a relatively large set of talented phrases as the candidates for the main components of the GLEAKE to explore the final keyphrases among them.

In order to achieve a high-quality set of final keyphrases, GLEAKE needs to extract a larger set of talented candidate phrases such that the coverage of the main topics of the input document is maximized. On the other hand, as Hulth investigated in [26], most manually assigned keyphrases follow the noun phrases pattern: $(NN.^* | JJ.^*)^* (NN.^*)$ where $NN.^*$ and $JJ.^*$ applies respectively to the noun tag and adjective tag with different morphologies. According to our research on many manually assigned keyphrases from different fields, the past participle form of a verb (VBN) can be considered as an adjective and the gerund form of a verb (VBG) can be used as a noun in composition of the keyphrases. Therefore, GLEAKE changes the noun phrase pattern to consider both VBN and VBG tags inside the pattern for the candidate phrase extraction task: $(NN.^* | JJ.^* | VBN | VBG)^* (NN.^* | VBG)$. After applying the customized noun phrase pattern and obtaining a large set of candidate phrases, GLEAKE then attempts to detect the outlier phrases and removes them from the set of candidates. According to our investigation, there is a direct relationship between the keyphrase length in terms of word count and its number of occurrences: the keyphrase with less words occurs more times than the keyphrase with more tokens. Therefore, GLEAKE employs a heuristic inequality to remove the outliers as shown in (1):

$$C < \alpha / ((n-1)^2 + 1) \qquad (1)$$

Where $C$ is the number of occurrences, $n$ is words count, and $\alpha$ is a hyperparameter. For each candidate, if the inequality is established, then the candidate will be removed from the set of candidate phrases as an outlier. According to our experiments, the best value of $\alpha$ hyperparameter for scientific and news contents is equal to seven. The final output of candidate extractor component is a set of clean talented phrases without any noisy data. This refined set of candidates is then used by two different mapper components to assign one local and one semantic vector to each phrase of the set.

### 3.2 Local Embedding Training and Mapping

After generating a set of candidate phrases, GLEAKE directly trains a word embedding model on only the input document and then employs it as the local mapper component to assign a structural vector for each candidate and the document. To do so, GLEAKE employs two well-known embeddings of skip-gram Word2Vec [27] and GloVe [28] to learn geometrical encodings of words from their co-occurrence information. As unsupervised techniques, both embeddings use neural networks to learn underlying word representations from an unlabeled textual corpus. On the other hand, according to our experiments, both of those techniques work very well with a small amount of training data such as a single textual document. The main difference is that the skip-gram is a "predictive" model, whereas GloVe is a "count-based" model. Otherwise stated, the skip-gram technique is designed to predict the context given a word, minimizing the loss of predicting the target context while GloVe aims to learn its vectors by basically performing dimensionality reduction on a word-word co-occurrence matrix by minimizing a reconstruction loss.

The word vectors generated by the local mapper capture the syntactical role of each word inside the input document. Considering this valuable property of word vectors, for each candidate phrase, GLEAKE simply sums the vectors of its words to obtain an analyzable local vector. On the other hand, for various domains, the beginning segment of a textual document contains the main concepts of the content. Therefore, GLEAKE selects first $M$ words of the input document as a representative and then sums their vectors into one vector as the local reference vector for that document. Our experiments indicate that the best value for the $M$ hyperparameter should be ten words where the representative contains the title and maybe a part of the beginning section of the input document. This local information is then integrated with semantical insight achieved from the global mapper in the next step to enable GLEAKE to construct an embedded-based graph.

### 3.3 Global Mapping Using (Multi) Word Embeddings

Although the local mapper captures the syntactic representations of the candidates and input document, GLEAKE also needs a semantic datasource to retrieve the keyphrases that are meaningfully related to the main topics of the input document. To provide such a comprehensive source for each specific domain, we have to train an efficient embedding model on huge amount of raw textual content. GLEAKE meets this requirement by leveraging various pre-trained single and multi-word embeddings from different domains of context. Table 1 depicts these pre-trained embedding models used by GLEAKE to assign a semantic vector for each candidate phrase. Among the investigated multi-word techniques in section 2.2, GLEAKE employs three models of fastText, Doc2vec, and Sent2vec to embed any sequence of words into a single fixed-length vector. Moreover, the proposed model utilizes global Word2Vec and GloVe models as single word embeddings to encode each word of a candidate phrase and then apply element-wise summing on the words' vectors to achieve a unified vector for that phrase. As shown in the Table 1, the embeddings are trained on large-scale corpora from various domains of scientific, news, common, medical, and social media. This property enables GLEAKE to choose a relevant embedding model and thus retrieve more semantically similar keyphrases to the content of the input document related to a specific area.

Similar to the previous step, GLEAKE selects $N$ beginning words of the input document as its representative and then assigns a global reference vector to this sequence of words. According to our experiments, the optimum value for $N$ is 250 beginning words that contain the title, beginning section, and maybe a part of the main content of the document.

### 3.4 Constructing Embedding-based Graph

All syntactic and semantic information obtained from two previous steps is merged into a single unit. In other words, in this step, GLEAKE constructs an embedding-based graph to integrate both local and global information into a unified component. In order to address the graph component more accurately, we aim to describe its construction using the mathematical terminology. For each phrase from the set of candidates $\{c_1, c_2, ..., c_n\}$ and also the input document $D$, there are two types of vectors: $(LV_{c_i}, GV_{c_i}), (LV_D, GV_D)$ where $LV$ and $GV$ respectively imply a local and global vector. Let $G = (V, E)$ specifies the Embedding-based Graph of input document where $\{c_1, c_2, ..., c_n\}$ and $E = \{(c_i, c_j) \mid i, j \in \{1, 2, ..., n\}\}$ is the set of edges between the candidates. GLEAKE uses eight different functions for weighting the edge of $(c_i, c_j)$ as listed in Table 2. It employs the cosine similarity to measure the similarity between two vectors of the same type. The first four methods only consider the vectors of $c_i$ and $c_j$ and calculate the local and global similarities between two candidates and finally combine the similarities by four different methods as shown in Table 2. How-ever, the next four methods also incorporate the document vectors into edge weighting composition. At first, they sum the vectors of the same type for $c_i$ and $c_j$, then measure the similarities between the resulting vectors and the document vectors, and ultimately integrate them using four different ways. All of the methods assign an edge to each pair of nodes if, and only if, both local and global cosine similarities are greater than zero, thus the $E$ set should be changed into $E = \{(c_i, c_j) \mid i, j \in \{1, 2, ..., n\} \text{ if } Local_{sim} > 0 \;\&\&\; Global_{sim} > 0\}$. As the results of this step, eight different embedding-based graphs are built and prepared for the graph mining operations in the final step in order to extract the final set of keyphrases.

TABLE 1
DIFFERENT PRE-TRAINED EMBEDDINGS USED BY GLEAKE

| Model Name | Source | Model | Vector Dimension | Model Reference |
|---|---|---|---|---|
| sent2vec_wiki_bigrams | WikiPedia | Sent2Vec | 700 | [22] |
| sent2vec_twitter_bigrams | Twitter | | | |
| pubmed2018_w2v_200D | PubMed | Word2Vec | 200 | [27] |
| wiki-news-300d-1M | WikiPedia + statmt.org news | FastText | 300 | [29] |
| doc2vec_wiki_dbow | WikiPedia | Doc2Vec | 300 | [30] |
| doc2vec_news_dbow | AP News | | | |
| glove.6B | Wikipedia + Gigaword | GloVe | 50-300 | [28] |
| glove.twitter.27B | Twitter | | 25-200 | |
| glove.840B | Common Crawl | | 300 | |

TABLE 2
DIFFERENT WEIGHTING FUNCTIONS FOR EMBEDDING-BASED GRAPHS

| Function Number | Method Composition |
|---|---|
| 1 | $F1(LV_{c_i}, LV_{c_j}) * F1(GV_{c_i}, GV_{c_j})$ |
| 2 | $F1(LV_{c_i}, LV_{c_j}) * F2(GV_{c_i}, GV_{c_j})$ |
| 3 | $F2(LV_{c_i}, LV_{c_j}) * F1(GV_{c_i}, GV_{c_j})$ |
| 4 | $F2(LV_{c_i}, LV_{c_j}) * F2(GV_{c_i}, GV_{c_j})$ |
| 5 | $F1(LV_{c_i} + LV_{c_j}, LV_D) * F1(GV_{c_i} + GV_{c_j}, GV_D)$ |
| 6 | $F1(LV_{c_i} + LV_{c_j}, LV_D) * F2(GV_{c_i} + GV_{c_j}, GV_D)$ |
| 7 | $F2(LV_{c_i} + LV_{c_j}, LV_D) * F1(GV_{c_i} + GV_{c_j}, GV_D)$ |
| 8 | $F2(LV_{c_i} + LV_{c_j}, LV_D) * F2(GV_{c_i} + GV_{c_j}, GV_D)$ |

Where $F1 = \cos(input_1, input_2)$ and $F2 = 1/(1 - \cos(input_1, input_2))$.

### 3.5 Score Assigning, Phrase Ranking, and Final Keyphrase Selecting

Until this step, GLEAKE has converted an input document into the embedding-based networks where the nodes imply the candidate phrases and the edges contain a combination of syntactic and semantic information. Considering such networks, there are different centrality measurements in network analysis to score each individual node based on its position and neighbors [31]. GLEAKE employs eight different centralities to assign numerical scores to the candidate phrases: Degree, eigenvector, page rank, personalized page rank, subgraph, harmonic, betweenness, and closeness. Let us focus on how the candidates get scores by applying each of the centralities on the embedding-based graph:

- Degree centrality counts the number of candidate phrases that the given phrase $c_i$ is connected to, normalized by the maximum possible degree in the graph.
- Closeness centrality of the phrase is calculated using (2), where $n$ is the number of candidates and $d(c_i, c_j)$ is the shortest-path distance between phrase $c_i$ and $c_j$:
  - $C_C(c_i) = (n-1) / (\sum_{j=1}^{n} d(c_i, c_j))$ (2)
- Betweenness centrality of the phrase $c_i$ is the sum of the fraction of all-pairs shortest paths that pass through $c_i$ as shown in (3).
  - $C_B(c_i) = \sum_{c_j, c_k \in V} (\sigma(c_j, c_k \mid c_i)) / (\sigma(c_j, c_k))$ (3)
- Harmonic centrality assigns a score to phrase $c_i$ according to (4), where $d(c_i, c_j)$ is the shortest-path distance between phrase $c_i$ and $c_j$ [32].
  - $C_H(c_i) = \sum_{i \neq j} (1/d(c_i, c_j))$ (4)
- Subgraph centrality for the phrase $c_i$ is the sum of weighted closed walks starting and ending at the phrase $c_i$ [33].
- Eigenvector centrality for the phrase $c_i$ is the i-th element of the vector $x$ defined by (5), where $A$ is the adjacency matrix of the embedding-based graph with eigenvalue $\lambda$.
  - $Ax = \lambda x$ (5)
- Page rank score of the phrase $c_i$ is calculated using (6), where $d$ is damping factor ranging from 0 to 1, $M(c_i)$ is the set of $c_i$'s neighbor phrases in the graph, $L(c_j)$ is the number of outbound edges from $c_j$, and $n$ is the number of candidates.
  - $C_{PR}(c_i) = (1-d)/n + d * \sum_{c_j \in M(c_i)} C_{PR}(c_j)/L(c_j)$ (6)
- Personalized page rank is a variation of page rank biased towards a set of phrases that their vectors are more similar to the document vectors. It assigns a score to the phrase $c_i$ using (7), where $W(c_i, D) = \cos(LV_{c_i}, LV_D) * \cos(GV_{c_i}, GV_D)$.
  - $C_{PPR}(c_i) = (1-d) * W(c_i, D)/n + d * \sum_{c_j \in M(c_i)} C_{PPR}(c_j)/L(c_j)$ (7)

After assigning the scores using each centrality measurement, GLEAKE sorts the candidate phrases based on their scores and selects top N of them as final keyphrases. In our experiment, we will separately investigate the accuracy measurements of different centralities.

## 4 EXPERIMENTS EVALUATION AND RESULT ANALYSIS

In this section, we will evaluate the performance of GLEAKE in terms of micro-averaged precision, recall, and F1-score measurements and demonstrate its improved performance in comparison to the five state-of-the-art and three baseline AKE methods. Three series of experiments are conducted: At first, we investigate the impact of four main hyperparameter s on the performance of GLEAKE. Then we analyze the local and global mappers of GLEAKE separately and indicate that a suitable combination of these two components leads to a set of high-quality keyphrases. Finally, we compare GLEAKE's efficiency with other AKE methods. Before concentrating on the experiments, we will describe five standard keyphrase datasets from news and scientific domains employed for our evaluation.

### 4.1 Datasets

GLEAKE has the capability to operate well on different domains and formats of text bodies. To illustrate this, we have employed five standard datasets from two different domains of news and scientific outlets where each dataset has a different style of writing and different length of words. The first dataset is the **SemEval-2010** dataset included 284 full-length scientific papers from the ACM digital library [6]. This set of papers is divided into three subsets: Trial, training, and test subsets that contain 40, 144, and 100 papers respectively. For each paper, there are three sets of standard keyphrases assigned by the human: author-assigned keyphrases (SemEval-A), reader-assigned keyphrases (SemEval-R), and a combined set of keyphrases from the SemEval-A and SemEval-R keyphrases (SemEval-C). For our evaluation, we use the test subset of articles and consider the SemEval-C keyphrases as a standard set of keyphrases.

The second dataset, provided by **Marujo** et al. [34], contains 500 news stories collected from the web. They employed multiple annotators to obtain several sets of keyphrases for each news story and then scored each phrase based on the number of annotators selecting the phrase as a keyphrase. The final set of keyphrases includes every keyphrase selected by more than 90 annotators. **Duc 2001** [35] is the other news dataset that we used to evaluate GLEAKE's performance. It contains 308 medium length newspaper articles from different newspaper resources which are categorized in 30 various topics.

**NUS**, as a scientific dataset, consists of 211 full conference papers where each paper contains between 4 and 12 pages [36]. Similar to the SemEval-2010, for each paper there are author-assigned and reader-assigned keyphrases, so we use both types of keyphrases for the GLEAKE evaluation. Since the final results of the GLEAKE evaluation on both SemEval-2010 and NUS are very similar, we use the SemEval-2010 in the "Hyperparameter Setting" and "Global and Local Embedding Analysis" sections and NUS in the "Comparison with State-of-the-Art Methods" section. The last dataset for evaluation is **Inspec**, a collection of 2000 short texts from scientific journal abstracts which was built by Hulth [26]. For each abstract, there are two kinds of keyphrases: controlled and uncontrolled. The controlled set is limited by a given dictionary while the uncontrolled ones are extracted by the experts. We only consider 500 testing abstracts for our evaluation with both controlled and uncontrolled keyphrases.

### 4.2 Hyperparameter Setting

Although the quality of the final set of keyphrases depends on various settings of GLEAKE's hyperparameters, different centrality measurements, edge weighting methods, local, and global embedding techniques are four key hyperparameters that strongly affect the performance of GLEAKE. In the following subsections, we are going to investigate each one with more analytical details to find its best configuration. To evaluate a certain parameter, the other parameter is set to the best value. Also, the number of keyphrases proposed by GLEAKE is set to $N=15$ in all experiments of this section and the next.

#### 4.2.1 Centrality Measurements

In order to achieve a set of comparable candidates, GLEAKE employs the centrality measurements to assign the importance score to each candidate phrase inside the embedding-based graphs. This score is determined with regard to the phrase place in the network topology so that different centralities use specific aspects of typical phrase connections with other candidates. Considering an embedding-based graph, since the weights of the edges are established based on the phrases' and the input document's vectors similarities, therefore the phrase with the higher score is more relevant to the main content of the input document. Table 3 shows the evaluation results for different centrality measures. According to these results, three centralities achieve the higher results among other measures: Page rank (PR), personalized PR, and eigenvector, with the best result yielded by eigenvector centrality. The interesting fact about these results is that both the page rank and personalized PR are variants of the eigenvector centrality where a node is important if it is linked to it by other important nodes. On the other hand, other well-known measures, such as betweenness and closeness yield a disappointing F1-score that shows their poor performance at the AKE field, although they are frequently used as high-efficiency metrics in other information retrieval and NLP tasks.

TABLE 3
PERFORMANCE OF CENTRALITY MEASUREMENTS

| Centrality | Precision | Recall | F1-score |
| --- | --- | --- | --- |
| Degree | 0.2019 | 0.2002 | 0.2010 |
| **Eigenvector** | **0.3270** | **0.3229** | **0.3249** |
| Page rank (PR) | 0.3215 | 0.3168 | 0.3191 |
| Personalized PR | 0.3240 | 0.3192 | 0.3216 |
| Subgraph | 0.1688 | 0.1682 | 0.1685 |
| Harmonic | 0.2011 | 0.1989 | 0.2000 |
| Betweenness | 0.0410 | 0.0421 | 0.0415 |
| Closeness | 0.1986 | 0.1947 | 0.1966 |

#### 4.2.2 Local Embedding Techniques

In order to capture the syntactic information, GLEAKE utilizes two embedding techniques of GloVe and Word2Vec to train a local word-vector mapper using only the content of the input document. The trained model has the capability to assign a local vector for each word of a typical phrase based on its occurrence position within the document. At a higher level, for each candidate phrase, GLEAKE sums the vectors of its words to obtain an analyzable syntactic vector. Although we have already discussed about the main functional differences between GloVe and Word2Vec embedding models, in practice, GloVe embeddings work better on some datasets, while word2Vec provides better performance on others. Table 4 demonstrates the superiority of the GloVe embeddings on the SemEval-2010 dataset, while Word2Vec model also achieves some comparable performance with the GloVe considering precision, recall, and F1-score metrics. For both embedding techniques, there are some internal hyperparameters discarded from analyzing in this paper. However, we consider the best configuration for each embedding model in GLEAKE's architecture.

TABLE 4
PERFORMANCE OF LOCAL EMBEDDINGS

| Model | Precision | Recall | F1-score |
|---|---|---|---|
| **GloVe** | **0.3270** | **0.3229** | **0.3249** |
| Word2Vec | 0.3041 | 0.3106 | 0.3073 |

### 4.2.3 Global Pre-Trained Models

One of the distinguishing features of GLEAKE is using an efficient global mapper to assign a semantic vector for each candidate phrase. To do so, our model employs two single and three multi-word embeddings trained on various domains of content as described in Table 1. The multi-word techniques assign a single fixed-length vector for any sequence of words while the unigram models are only able to map one word to a fixed-length vector where a semantic encoding for each candidate phrase is obtained by summing its words' vectors. Table 5 shows the influence of the different global mappers on the quality of final keyphrases considering precision, recall, and F1-score measurements. As expected, for scientific-based embeddings, GLEAKE achieves higher performance than non-scientific ones and the best result belongs to the DOC2Vec model trained on the wikipedia corpus. On the other hand, multi-word embeddings produce better results in comparison to the single word embeddings. This comparison indicates that tracking a sequence of words as a unified token during the training process captures more semantic insights about the keyphrases than considering words as single tokens.

TABLE 5
PERFORMANCE OF GLOBAL EMBEDDINGS

| Model Name | Precision | Recall | F1-score |
|---|---|---|---|
| sent2vec_wiki_bigrams | 0.3188 | 0.3182 | 0.3185 |
| sent2vec_twitter_bigrams | 0.2682 | 0.2760 | 0.2720 |
| pubmed2018_w2v_200D | 0.1957 | 0.2325 | 0.2125 |
| wiki-news-300d-1M | 0.3041 | 0.2964 | 0.3002 |
| **doc2vec_wiki_dbow** | **0.3270** | **0.3229** | **0.3249** |
| doc2vec_news_dbow | 0.2451 | 0.2600 | 0.2523 |
| glove.6B | 0.2308 | 0.2732 | 0.2502 |
| glove.twitter.27B | 0.1575 | 0.1829 | 0.1692 |
| glove.840B | 0.1950 | 0.2121 | 0.2032 |

### 4.2.4 Edge Weighting Functions

GLEAKE uses eight heuristic functions to assign a weight for each edge between two candidate phrases inside the embedding-based graph. Each function applies different way for integrating the local and global features into one scalar number indicating the edge value. Table 6 Indicates the evaluation results for these functions where the GloVe, doc2vec_wiki_dbow, and eigenvector centrality are employed for the local embedding, global embedding, and candidates scoring respectively. As can be seen, the first four functions obtain low accuracy considering precision, recall, and F1-score measurements, these methods only consider the embedded vectors of candidate phrases in their calculation. On the other hand, the last four functions also consider the embedded vector of the input document into their compositions where the semantic and syntactic similarities between the document and candidates determine the weights of the edges, thus their remarkably better performance implies to a high-quality set of final keyphrases. For the SemEval-2010 dataset, the sixth combination method achieves the best performance indicating the global similarity scores between the document and candidates are denser than the local similarities.

TABLE 6
THE ACCURACY RESULTS FOR DIFFERENT WEIGHTING FUNCTIONS

| Function Number | Precision | Recall | F1-score |
|---|---|---|---|
| 1 | 0.1729 | 0.1716 | 0.1722 |
| 2 | 0.1862 | 0.1830 | 0.1845 |
| 3 | 0.1250 | 0.1253 | 0.1251 |
| 4 | 0.1244 | 0.1200 | 0.1221 |
| 5 | 0.3215 | 0.3183 | 0.3199 |
| **6** | **0.3270** | **0.3229** | **0.3249** |
| 7 | 0.2891 | 0.2872 | 0.2881 |
| 8 | 0.2823 | 0.2800 | 0.2811 |

## 4.3 The Local and Global Embedding Analysis

The embedding components of GLEAKE are able to operate independently and produce some high-quality keyphrases separately. However, each component only considers one significant property of the candidate phrase to select the best set among them. In this section, we concentrate on analyzing the global and local mappers of GLEAKE individually and demonstrate how their properly integrated model, GLEAKE, outperform both of the mappers on the scientific content.

Table 7 and Table 8 show the standard keyphrases for the C-30 and H-48 SemEval articles alongside the extracted keyphrases using the local (L), global (G), and the combined model (C) for those articles. The numbers inside the cells imply the rank of the extracted keyphrase by the model indicated in the related column header. For C-30 article, each of the local and global models finds four original keyphrases among the first fifteen extracted keyphrases where two found keyphrases using them are the same. An analytical look at the order of the locally extracted keyphrases indicates the structural importance of the input document, while the arrangement of the keyphrases by the global model reflect the semantical emphasis of the keyphrases. Using a suitable combination, GLEAKE successfully covers all of the found keyphrases by the global and local models among the top fifteen found keyphrases as shown in the third column of Table 7. The same analysis can also be used for the H-48 article as shown in Table 8.

TABLE 7
A SET OF STANDARD KEYPHRASES FOR THE C-30 PAPER IN SEMEVAL-2010.

| Keyphrase (C-30:Doc2Vec) | L | G | C |
|---|---|---|---|
| overlay mesh | 13 | 15 | 14 |
| data dissemination | 1 | - | 4 |
| overlay network | - | 5 | 6 |
| ip multicast | - | - | - |
| multipoint communication | - | - | - |
| high-bandwidth data distribution | - | - | - |
| large-file transfer | - | - | - |
| real-time multimedia streaming | - | - | - |
| bullet | 2 | 2 | 2 |
| bandwidth probing | - | - | - |
| peer-to-peer | - | - | - |
| ransub | - | - | - |
| content delivery | - | - | - |
| tfrc | - | - | - |
| bandwidth | 3 | - | 3 |
| overlay | - | 9 | 10 |
| Number of found keyphrases | 4 | 4 | 6 |

*As can be seen, GLEAKE has found six of sixteen standard keyphrases among its top fifteen extracted keyphrases, indicated by the last row and thus achieves better performance in comparison to local and global mappers. Each number shows the rank number of the found keyphrase and the '-' sign implies the keyphrase is not found by the related method specified at the column's head.*

TABLE 8
A SET OF STANDARD KEYPHRASES FOR THE H-48 PAPER IN SEMEVAL-2010.

| Keyphrase (H-48:Sent2Vec) | L | G | C |
|---|---|---|---|
| query expansion | 1 | 3 | 2 |
| query-document term mismatch | - | 1 | 5 |
| information retrieval | 2 | 15 | 9 |
| information search | - | - | - |
| document expansion | - | - | - |
| document processing | - | - | - |
| relevant document | 7 | - | 6 |
| evaluation | - | - | 10 |
| Number of found keyphrases | 3 | 3 | 5 |

*GLEAKE has found five of eight standard keyphrases among its top fifteen extracted keyphrases, indicated by the last row and thus achieves better performance in comparison to local and global mappers. Each number shows the rank number of the found keyphrase and the '-' sign implies the keyphrase is not found by the related method specified at the column's head.*

### 4.4 Comparison with State-of-the-Art Methods

In sections 4.2.1 and 4.2.2, we investigated the influence of two main hyperparameters on the GLEAKE efficiency and determined that the best result is yielded by the configuration of [centrality: Eigenvector, combination method ID: 6, global embedding model: Doc2Vec300DBOW_wiki, local embedding model: glove] when we employ the SemEval-2010 dataset for our evaluation. In this section, we aim to compare GLEAKE's performance to five state-of-the-art AKE methods: CopyRNN [18], EmbedRank [37], TSAKE [38], WikiRank [39], Ying [40]. To do so, we utilize four standard datasets from two different domains of news and scientific outlets as explained earlier: Marujo, DOC 2001, NUS, Inspec. Before focusing on comparison, Table 9 shows the best configuration of GLEAKE on the four standard AKE datasets obtained using experimentation of different configurations as outlined in the previous sections. For all datasets, the Personalized PageRank and Eigenvector metrics have higher efficiency among the other centrality measures indicating the importance of a phrase based on the number and powerfulness of its neighbors. On the other hand, the best combination IDs are related to the last four combination methods where the document vector is also incorporated into edge weighting composition. However, for two other hyperparameters, there is no specific solution that dominates the other related alternatives among the available solutions. This issue shows GLEAKE flexibility when it confronts with new standard dataset from different domains and styles as it can investigate the AKE task among various options of practical combinations and track the best configuration for every specific dataset.

TABLE 9
BEST CONFIGURATIONS OF THE HYPERPARAMETER S FOR STANDARD AKE DATASETS

| Dataset | Centrality | Function Number | Global Model | Local Model |
|---|---|---|---|---|
| DOC 2001 | Personalized PR | 2 | wiki-news-300d-1M.vec | word2vec |
| Marujo | Eigenvector | 6 | Doc2Vec300DBOW_news | glove |
| NUS | Personalized PR | 7 | pubmed2018_w2v_200D | word2vec |
| Inspec | Personalized PR | 5 | Sen2Vec700wiki | glove |

We have employed the mentioned configurations in Table 9 to adjust GLEAKE's hyperparameters and then compared its performance with other AKE methods. For NUS, Inspec, and DOC 2001 datasets, we set the number of the final keyphrases equal to $N = 10$ in order to achieve comparable results with the other AKE methods under investigation. This value for the Marujo dataset is equal to $N = 10$. Table 10 and Table 11 present the results of performance comparison in terms of micro-averaged precision, recall and F-measure metrics. According to the performance results, GLEAKE outperforms most of state-of-the-art AKE methods in almost all of the standard datasets. The only exception is the EmbedRank method that yields better results on the Duc 2001 dataset. However, this method presents poor performance on the NUS dataset and also has lower efficiency on the Inspec dataset in comparison to our method. According to our wide investigation on many modern AKE methods, most of them do not provide comprehensive experiments on standard AKE datasets from different domains and styles. For the rare cases of doing so, the final performance is considerably poor in at least one of the datasets under investigation. Table 10 and Table 11 not only verify this claim but also demonstrate GLEAKE could successfully achieve a high-quality set of final keyphrases on all different dataset with various domains and styles.

Most of the investigated AKE methods in this section have used structural or semantic properties of the candidate phrases to find a set of the final keyphrases. CopyRNN [18] captures semantic and syntactic insights using recurrent neural networks to compress the meaningful information, at the same time applying a copying mechanism to find important parts based on positional information. Motivated by the fact that a word must be important if it appears in many relevant sentences, Ying et

TABLE 10
COMPARISON OF GLEAKE'S PERFORMANCE WITH OTHER AKE METHODS ON NUS AND MARUJI DATASETS

| AKE Method | NUS | | | Marujo | | |
|---|---|---|---|---|---|---|
| | P | R | F | P | R | F |
| EmbedRank [37] | 0.1368 | 0.1394 | 0.1381 | - | - | - |
| TSAKE [38] | - | - | - | 0.1430 | 0.4660 | 0.2190 |
| CopyRNN [18] | - | - | 0.3170 | - | - | - |
| WikiRank [39] | 0.0727 | 0.1216 | 0.0910 | - | - | - |
| Ying [40] | - | - | - | **0.4870** | 0.4980 | 0.4780 |
| GLEAKE | **0.3469** | **36.63** | **0.3563** | 0.4565 | **0.5054** | **0.4797** |

*The '-' sign indicates that no performance statistics is provided by a method for a specific dataset.*

TABLE 11
COMPARISON OF GLEAKE'S PERFORMANCE WITH OTHER AKE METHODS ON INSPEC AND DUC DATASETS.

| AKE Method | Inspec | | | Duc | | |
|---|---|---|---|---|---|---|
| | P | R | F | P | R | F |
| EmbedRank [37] | 0.3575 | 0.4040 | 0.3794 | **0.2882** | **0.3558** | **0.3185** |
| TSAKE [38] | 0.4010 | 0.2030 | 0.2690 | - | - | - |
| CopyRNN [18] | - | - | 0.3360 | - | - | - |
| WikiRank [39] | 0.2814 | 0.2597 | 0.2701 | 0.2872 | 0.2644 | 0.2753 |
| Ying [40] | 0.4300 | 0.4020 | 0.3960 | - | - | - |
| GLEAKE | **0.4804** | **0.3645** | **0.4145** | 0.2538 | 0.3256 | 0.2853 |

*The '-' sign indicates that no performance statistics is provided by a method for a specific dataset.*

al. [40] apply an iterative ranking algorithm on a network with three levels of structural relationships to rate the words of the input document, and then leverage clustering technique to obtain several clusters as the document's main topics and finally choose the candidate phrase close to the centroid of each cluster as final keyphrase. EmbedRank [37] uses two well-known pre-trained embedding models to automatically extract the keyphrases. The model guarantees two important properties of keyphrases: 1.) informativeness by using the cosine similarity between the keyphrases and document embedding vectors, and 2.) diversity to combine keyphrase informativeness with dissimilarity among the selected keyphrases. The WikiRank [39] constructs a semantic graph from the input document including topical concepts of Wikipedia, and then prunes the graph to filter out candidates which are likely to be erroneously generated. This approach generates a final set of keyphrases that has optimal coverage of the main concepts of the document. As an unsupervised AKE method, TSAKE [38] generates several topical graphs by assigning topical weights to the edges of the document word graph and then applies network analysis techniques to each topical graph to detect finer grained topics. This approach uses the central words of the minor topics to rank and select top N candidates as the final keyphrases.

Despite the smart approaches employed by the state-of-the-art methods, two outstanding features of GLEAKE reveal its superiority over those intelligent methods: At first, GLEAKE has the capability to employ a proper global embedding model for exploring the semantic clues based on the content type of the input document. Therefore, unlike some recent keyphrase extraction approaches [18], [38], GLEAKE is free from any specific domains of knowledge and so is able to find a set of high-quality keyphrases in different domains such as scientific outlets, news, and novels. The GLEAKE performance on two scientific and news datasets (Table 10 and Table 11) illustrates this as our model could successfully overcome most of the state-of-the-art AKE methods. The second distinguishing feature of GLEAKE is its comprehensive view of predicting each token of a typical keyphrase: some AKE methods [37], [39] consider only the semantic aspect of each token without paying attention to syntactic features of it inside the source text, while some others [40] focus on local features of the candidates without considering the semantic aspect of them. However, GLEAKE employs modern (multi)-word embedding techniques to automatically draw the syntactic and semantic features out of each candidate keyphrase and utilizes a scientific technique based on network analysis to find an optimal set of final keyphrases based on both local and global information.

We completed this section by comparing GLEAKE's performance with three baseline methods in terms of precision-recall diagrams: TFIDF, LDA-based, Centrality-based. As a classical statistic in the NLP area, the TFIDF computes the importance of a word inside an input document by integrating its term frequency ($tf$) in the document with its inverse document frequency ($df$) from a large document corpus: $TFIDF_w = tf_w * \log(N / df_w)$, where $N$ is the number of documents within the corpus under investigation. For each candidate phrase, there are two ways of TFIDF computation: summing the scores for the words of the phrase or averaging those scores. For our comparison analysis, we consider the best performance between them as TFIDF results. The LDA-based method computes the topical (semantical) similarity between each candidate phrase and the input document to score the candidates. To do so, a pre-trained LDA model [14] is employed to assign a distribution vector of different topics for phrase $p: \Phi(T, p)$ and document $D: \Phi(T, D)$, where $T$ is a set of different topics. Finally, the cosine similarity is applied on the topic distributions of each candidate and the document to calculate a semantic similarity for that phrase as its final score. The Centrality-based model builds an occurrence graph with nodes as candidates and edges as their co-occurrence counts and then uses different centrality measures to assign scores to the nodes and finally selects the candidates with the highest scores as final keyphrases set.

Fig. 2 depicts various precision-recall diagrams on different AKE datasets investigated in this paper. Each spot located on the diagrams is related to a specific number of final keyphrases ($N$), so that every full diagram contains the results for different numbers of final keyphrases in the range of $N = 5, 6, ..., 15$. As expected, our approach considerably outperforms all baselines on different $N$s and datasets. These results clearly demonstrate irrefutable superiority of (multi)-word embedding techniques compared to different classical models. Among these three baseline methods, the Centrality-based model presents the better F-measure result implying the importance of central nodes in extracting a final set of keyphrases. On the other hand, the TFIDF method provides the lowest performance as it only uses simple syntactic statistics without considering the semantic role of words or phrases inside the document.

## 5 CONCLUSION AND FUTURE WORK

In this paper, we proposed GLEAKE, an unsupervised method for automatic keyphrase extraction using integration of syntactical and semantical insights. GLEAKE believes that each real keyphrase has two essential properties that characterize its identity: the first feature is related to the local role of the keyphrase in a specific context and the second feature is the general concept of the keyphrase. To address these properties, GLEAKE employs phrase embedding techniques to discover the local and global attributes of the candidate phrases and then merges them into a series of embedding-based graphs. GLEAKE utilizes network analysis techniques to rank the candidate keyphrases. We conducted three series of experiments to evaluate the performance of GLEAKE on scientific and news domains. The state-of-the-art results clearly confirm the efficiency of our proposed method for extracting high-quality keyphrases from different domains and styles. As our future work, we aim to investigate the practicality of our approach in other domains of textual contents and also adjust it to other relevant tasks, such as document summarization and categorization.

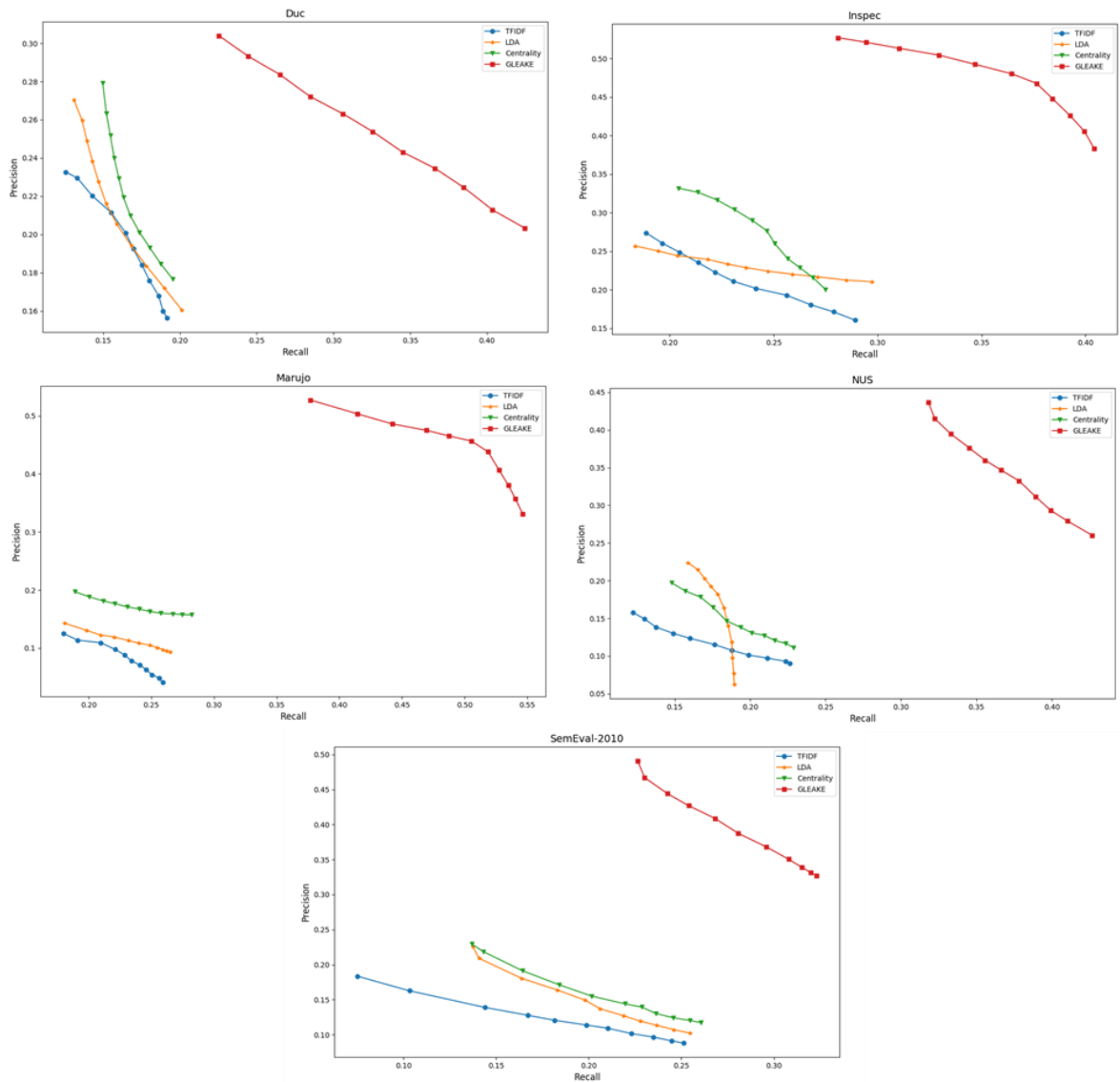

Fig. 2. Precision and recall diagrams for the baseline methods compared with the GLEAKE

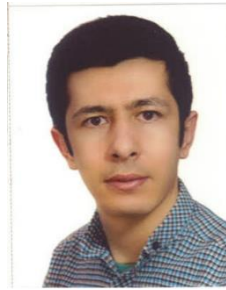

**Javad Rafiei Asl** recived the BS degree from the University of Tabriz in 2011 and the MS degree from the University of Tehran in 2014. He is a PhD student at Computer Sceince Department of Georgia State University. His current research interests are natural language processing (NLP) and desinging new deep learning techniques to solve different NLP tasks considering large-sacle and middle-scale datasets.

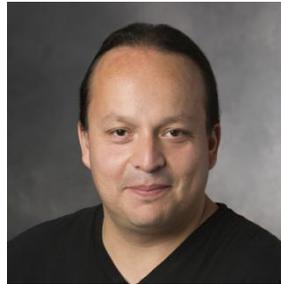

**Juan M. Banda** is currently an assistant professor of computer science at Georgia State University. He holds a PhD and MS in computer science and mathematics from respectively the Montana State University and Eastern New Mexico University. Prior to joining Georgia State University, Juan was a Postdoctoral Data Science Fellow at Shah Lab where he looked to uncover information hidden in the dark corners of the free-text sections of the EMR systems of Stanford's STRIDE data warehouse. As an engineer at heart and practice for the last 20 years, he has used Python, Bash, ontologies, and NLP tools to build pipelines to annotate over 68 million clinical notes.